# Gallium Nitride FET Model

V V Orlov, G I Zebrev

National Research Nuclear University MEPHI, Moscow, Russia

E-mail: gizebrev@mephi.ru

**Abstract.** We have presented an analytical physics-based compact model of GaN power FET, which can accurately describe the I-V characteristics in all operation modes. The model considers the source-drain resistance, different interface trap densities and self-heating effects.

## 1. Introduction

Gallium nitride (GaN) high electron mobility transistor (HEMT) technology has many advantages, that make it a promising candidate for high-speed power electronics. It allows high-power operation at much higher frequencies than silicon laterally diffused metal-oxide-semiconductor field-effect transistors (LDMOSFETs), currently a staple for the cellular base station industry [1]. The high breakdown voltage capability (over 100 V), high electron mobility, and high-temperature performance of GaN HEMTs are the main factors for its use in power electronics applications. Circuits design in both application regimes requires the accurate compact device models that can describe the non-linear I-V characteristics. The current state-of-the-art GaN power transistor circuit models are mostly empirical in nature and contain a large number of fitting parameters. The source-drain series resistance and self-heating make the compact modeling difficult [2]. Currently available models are not enough accurate to describe the I-V characteristics of power GaN HEMTs in all operation modes. This means, that we need a compact physics-based analytical model based on the physical description of the device. In this paper, we present a physics-based GaN power transistor model based on generic approach The paper contains 3 parts. In the first part, we will give a concise description of the model. The specific power HEMT's effects, such as series resistance and self-heating will be discussed in the second and third parts.

## 2. Compact model Gallium Nitride FET

*2.1. Concise description of the compact model*

Presented in this paper model relies on the diffusion-drift transistor model, originally proposed in [3,4]. This model is based on an explicit solution of the channel current continuity equation, and it has been consistently realized for different types field-effect devices including SOI and double-gate transistors [5], graphene FETs [6], and molybdenite $MoS_2$ monolayer transistor [7]. This model is able to describe via a single analytic expression the whole I-V characteristics both in subthreshold and above threshold regions and in linear and saturation modes.

A generic form I-V characteristics of an ideal (i.e., without any geometrical short-channel effects) FET in a concise form

$$I_D = I_{DSAT}\left(1 - \exp\left(-2\frac{V_{DS}}{V_{DSAT}}\right)\right). \tag{1}$$

where $I_{DSAT}$ is the saturation current, $V_{DS}$ is the drain-source bias. The drain saturation voltage $V_{DSAT}$ is an explicit function of the mobile charge density

$$V_{DSAT} \cong \varphi_T\left(1 + \frac{C_{it}}{C_{ox} + C_D}\right) + \frac{qn_{S0}}{C_{ox} + C_D}, \tag{2}$$

where $q$ is the electron charge, $n_{S0}$ is the channel carrier density near the source, $\varphi_T = k_B T/q$ is the thermal potential, $C_{ox}$, $C_D$ are the oxide and the depletion layer capacitance per unit area, correspondingly. The MOSFET electrostatic saturation current $I_{DSAT}$ is represented as a sum of the diffusion and the drift components [8].

$$I_{DSAT} = \frac{W}{L} q D_0 n_{S0} + \frac{W}{L} \frac{\mu_0 q^2 n_{S0}^2}{2(C_{ox} + C_D)}, \tag{3}$$

where $W$ and $L$ are the channel's width and length, $q$ is the electron charge, $\mu_0$ and $D_0$ are the channel carrier mobility and diffusivity, coupled by the Einstein relation $D_0 = \mu_0 \varphi_T$. The first term in (3), corresponding to a linear dependence of the drain current on $n_{S0}$, is a diffusion current, dominating in the subthreshold region at low $n_{S0}$. The second term in (3) corresponds to the saturated drift current in the above-threshold mode in a well-known square-law approximation [9]. The dependence on the gate voltage $V_G$ has in (1) an implicit form via the dependence $n_{S0}(V_G)$. A concrete form of $n_{S0}(V_G)$ should be determined by electrostatic consideration, which is different for different device configurations (bulk or SOI FETs, double gate FETs, FinFETs etc.)

$$I_D = \frac{W}{L} e\mu_0 n_{S0} V_{DS} \equiv G_D V_{DS}, \tag{4}$$

where $G_D$ is the channel conductance (i.e. total inverse channel resistance at low drain bias) in a low drain bias regime

$$G_D = \frac{W}{L} e\mu_0 n_{S0}. \tag{5}$$

*2.2. Channel carrier density as function of gate voltage*

We will use a phenomenological interpolation for $n_{S0}(V_G)$ which is similar to that used in CMOS design compact model BSIM [10]

$$en_{S0} = \frac{2C_1 S \ln\left[1 + \exp\left(\frac{V_G - V_T}{2S}\right)\right]}{1 + 2\frac{C_1 S}{C_2 \varphi_T} \exp\left(-\frac{V_G - V_T}{2S}\right)}, \tag{6}$$

where $V_T$ is the threshold voltage, $C_1$ ($C_2$) is the effective gate-to-channel capacitance per unit area in the above threshold (subthreshold) operation modes, $S = dV_G/d(\ln I_D)$ is the logarithmic subthreshold slope [9] which to be fitted from the sub-threshold part of the I-V curve. The interpolation (6) is validated by its asymptotic in the subthreshold and above-threshold regions (note $SS \equiv \ln 10\, S$)

$$en_{S0} \cong \begin{cases} C_1(V_G - V_T), & V_G > V_T, \\ C_2 \varphi_T \exp\left(\dfrac{V_G - V_T}{S}\right) = C_2 \varphi_T 10^{\frac{V_G - V_T}{SS}}, & V_G < V_T. \end{cases} \qquad (7)$$

In GaN FET the subthreshold slope is strongly dependent on the drain voltage

$$S = S_0 + m_d V_{DS}, \qquad (8)$$

where $S_0$ (~ 90-120 mV/decade) is the constant subthreshold slope extracted at small $V_{DS}$, $m_d$ (~0.01-0.2) is a fitting constant. The threshold voltage is dependent on $V_{DS}$ via the so-called DIBL (Drain-Induced Barrier Lowering) effect

$$V_T = V_{T0} - \alpha_{DIBL} V_{DS}, \qquad (9)$$

where $V_{T0}$ is a constant threshold voltage, $\alpha_{DIBL}$ is a DIBL sensitivity parameter.

## 3. Accounting for series resistance

Most of the state-of-the-art power FETs are not self-aligned and their access regions affect device performance. The external drain voltage $V_D$ can be written as a sum of internal channel voltage $V_D^{int}$ and the voltage on $R_D$

$$V_D^{int} = V_D - I_D R_D, \qquad (10)$$

where $R_D$ is the drain resistance. Taking into account the equation (1), one can rewrite (10) as follows

$$\frac{V_D - V_D^{int}}{R_D} = I_{DSAT}(V_G)\left(1 - \exp\left(-2\frac{V_{DS}^{int}}{V_{DSAT}(V_G)}\right)\right), \qquad (11)$$

where

$$I_{DSAT} = \frac{1}{2} G_D V_{DSAT}. \qquad (13)$$

From the equation (11) we have the exact solution for the drain current

$$I_D \cong I_{DSAT}\left(1 - \frac{1}{s} W\left[s \exp\left(s - \frac{2V_{DS}}{V_{DSAT}}\right)\right]\right), \qquad (14)$$

where $s = G_D R_D$, $W(s)$ is the Lambert function, defined as a solution of the equation

$$W(se^s) = s. \qquad (15)$$

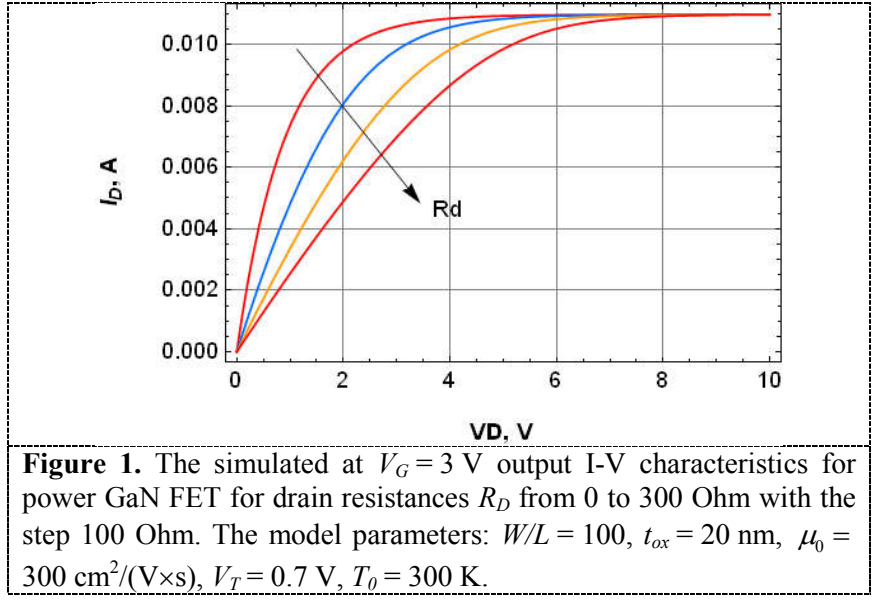

**Figure 1.** The simulated at $V_G = 3$ V output I-V characteristics for power GaN FET for drain resistances $R_D$ from 0 to 300 Ohm with the step 100 Ohm. The model parameters: $W/L = 100$, $t_{ox} = 20$ nm, $\mu_0 = 300$ cm$^2$/(V×s), $V_T = 0.7$ V, $T_0 = 300$ K.

Figure 1 shows the output current-voltage characteristics for power GaN FET. Note, that the drain current decreases with the increasing of drain resistance $R_D$.

## 4. Self-heating effects

The problem of self-heating in power GaN HEMTs is a severe issue that that affects the device reliability and degrades the I-V characteristics [11]. The carrier mobility at room and elevated temperatures T is mainly determined by phonons, and it can be simulated as to be proportional to $T^{-n}$ where $n$ is a parameter (~1.0-1.5). Mobility is generally reducing the temperature due to phonon scattering

$$\mu_0(T) \propto \left(\frac{T_0}{T_0 + \Delta T}\right)^n \propto 1 - n\frac{\Delta T}{T_0}, \tag{16}$$

where $T_0$ is a nominal operating temperature. We will model the self-heating effects through the following self-consistent equation

$$I_D(T_0 + \Delta T)V_{DS}R_{th} = \Delta T, \tag{17}$$

where $R_{th}$ is the thermal resistance (~ 30 K/W). From the equation (17) one gets

$$I_D(T_0)\left(1 - n\frac{\Delta T}{T_0}\right) \cong \frac{\Delta T}{V_{DS}R_{th}}, \tag{18}$$

$$\Delta T \cong I_D(T_0)\frac{V_D R_{th}}{nI_D(T_0)V_{DS}R_{th} + 1}. \tag{19}$$

Here, $n$ can be considered as a fitting dimensionless parameter of order unity. Substituting (19) to the equation (17), the drain current with self-heating effects $\tilde{I}_D$ can be approximated by the following relationship

$$\tilde{I}_D = I_D(T_0 + \Delta T) \cong I_D(T_0) - n\frac{I_D(T_0)}{T_0}\Delta T = \frac{I_D(T_0)}{1 + nI_D(T_0)V_{DS}R_{th}/T_0}. \tag{20}$$

Figure 2 shows the dependencies of the drain currents $I_D$ as functions of drain voltages $V_D$ at different carrier mobility with the self-heating effect. The drain current decreases with the increasing of drain voltage in the saturation mode.

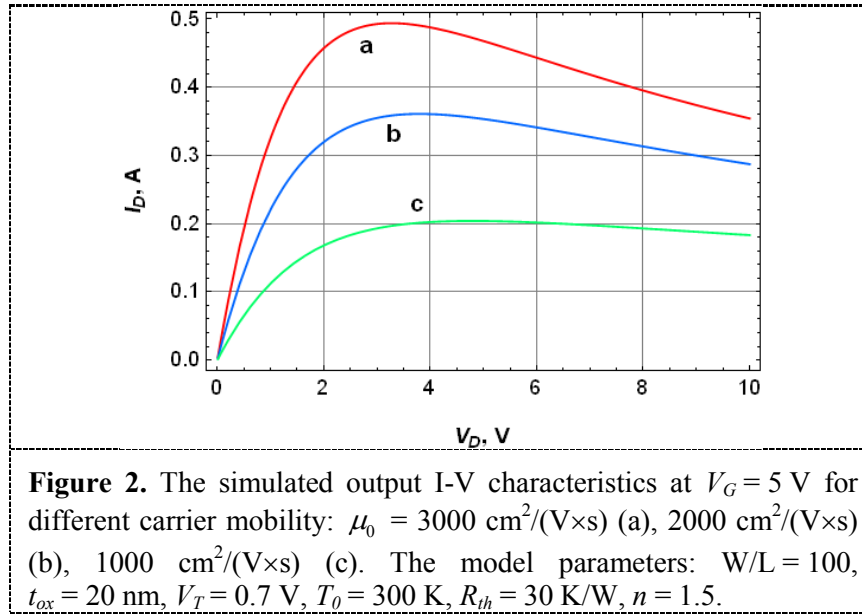

**Figure 2.** The simulated output I-V characteristics at $V_G = 5$ V for different carrier mobility: $\mu_0 = 3000$ cm$^2$/(V×s) (a), 2000 cm$^2$/(V×s) (b), 1000 cm$^2$/(V×s) (c). The model parameters: W/L = 100, $t_{ox} = 20$ nm, $V_T = 0.7$ V, $T_0 = 300$ K, $R_{th} = 30$ K/W, $n = 1.5$.

## 5. Conclusions

Our goal has been to provide a description of the power GaN HEMT compact analytical model, which is capable of simulating in a unified manner all the FET's operation modes using a compact and transparent analytical relationship. It has been shown that the model is suitable for accurate simulation of the GaN I-V characteristics in modern power HEMT circuits. The simulation results show that the analytical model is useful for describing the I-V characteristics, taking into account the drain resistance and self-heating effects.